\documentclass[aps,pre,twocolumn,superscriptaddress,showpacs]{revtex4}
\usepackage{amsfonts,amssymb,amsmath,latexsym,epsfig} 
\begin{document}

\title{Irreversible growth of binary mixtures on small-world networks}

\author{Juli\'an Candia}
\affiliation{{The Abdus Salam International Centre for Theoretical Physics,}\\ 
{Strada Costiera 11, 34014 Trieste, Italy}}

\begin{abstract}
Binary mixtures growing on small-world networks under far-from-equilibrium conditions are 
studied by means of extensive Monte Carlo simulations. For any positive value of the 
shortcut fraction of the network ($p>0$), the system undergoes a continuous order-disorder 
phase transition, while it is noncritical in the regular lattice limit ($p=0$). 
Using finite-size scaling relations, the phase diagram is obtained in the thermodynamic limit  
and the critical exponents are evaluated. 
The small-world networks are thus shown to trigger criticality, a phenomenon analogous to 
similar observations reported recently in the investigation of equilibrium systems. 
\end{abstract}


\maketitle

\section{Introduction}

Complex networks are known to play a key role in the description of the structure 
and evolution of different mesoscopic and macroscopic systems.
Indeed, the interacting parts of many natural and artificial systems can be interpreted 
as collections of linked nodes forming complex networks, whose structure and topology can be characterized 
in terms of statistical quantities such as their degree and path-length distributions, their connectivity, etc. 

Empirical observations show that the 
mean distance between a pair of nodes within a connected real network is in general surprisingly short
(typically of a few degrees and only logarithmically dependent on the system size), a 
phenomenon known as {\it small-world effect}. Moreover, the neighborhood of each node is 
observed to be, on average, highly interconnected \cite{dor03}. 

The well-studied classical random graphs, which are networks built by linking nodes at random, 
display the small-world effect but have much lower connectivities than usually observed in real networks.   
In this context, the small-world networks were proposed few years ago \cite{wat98, wat99} 
as a realization of complex networks having short mean path-lengths 
(and hence showing the small-world effect) as well as large connectivities. 
Starting from a regular lattice, a small-world network is built by randomly adding or 
rewiring a fraction $p$ of the initial number of links. Even a small fraction of added or rewired links 
provides the shortcuts needed to produce the small-world effect, thus displaying a global behavior close to 
that of a random graph, 
while preserving locally the ordered, highly connected structure of a regular lattice. 
Indeed, it has been shown that this small-world regime is reached for any given disorder probability $p > 0$, 
provided only that the system size $N$ is large enough (i.e. $N>N_c$, 
where the critical system size is $N_c\propto 1/p$) \cite{new99a, bar00}.

As a further step, recent works investigated the behavior of many standard models of Statistical Mechanics 
defined on small-world networks (see e.g. \cite{bar00,git00,kim01,hon02,med03,hin05}), as well as on other classes of 
complex networks \cite{dor03}. 
In particular, this was done for several equilibrium, Ising-type spin models. 
Besides their inherent theoretical interest, spin models defined on small-world networks can describe 
structural properties of polymer chains \cite{mon00,jes00,sca01} and have also been found useful in 
the study of social phenomena.
For instance, spin states may denote different opinions or preferences, where the  
coupling constant describes the convincing power between interacting individuals, 
which is in competition with the ``free will'' given by the thermal noise \cite{ale02}. Moreover, 
a magnetic field can be used to add a bias that could be interpreted as ``prejudice'' or ``stubbornness'' \cite{sve02}. 

Generally speaking, it was found that the structure and topology of the underlying complex networks 
affect dramatically the critical behavior of the models defined on them. For instance, it was found 
that the Ising model defined on a 1D small-world network presents a second-order phase transition 
at a finite critical temperature $T_c$ for any value of the rewiring probability $p>0$ \cite{bar00, git00}. 
Considering directed links, even the nature of the phase transition was found to change, switching from second to 
first order \cite{san02}. 
The ferromagnetic transition for the Ising model on 
small-world networks has also been studied numerically by rewiring 2D and 3D regular lattices \cite{her02}. 

Much less attention, however, has been devoted so far to the investigation of nonequilibrium transitions 
on complex networks. Some simple nonequilibrium models closely related to percolation were initially 
studied \cite{kup01,pas01,egu02}, while more recently a model for social interaction was investigated \cite{kle03},
in which the competition between dominance and spatial coexistence of different states in the nonequilibrium 
dynamics of Potts-like models was examined. 
Within the context of these recent developments, the aim of this work is to investigate the irreversible growth 
of binary mixtures on small-world networks. 

The growth of a binary mixture (or, adopting an equivalent magnetic language, a two-state magnetic system 
of up and down spins) can be studied by means of the so-called magnetic Eden model 
(MEM) \cite{van94,can01}, a natural generalization of the classical Eden model \cite{ede58} 
in which the particles have an additional degree of freedom, the spin. 
In regular lattices, the MEM's growth process leads to an Eden-like self-affine growing interface and 
a fractal cluster structure in the bulk, and displays a rich variety of 
nonequilibrium phenomena, such as thermal order-disorder continuous phase transitions,  
spontaneous magnetization reversals, as well as morphological, wetting, and corner wetting transitions.
While the Eden model has  
extensively been applied to many different problems (namely, soot formation \cite{her86}, colloids \cite{her86}, 
percolation \cite{bun85}, growth of cell colonies \cite{bar94}, crystal growth \cite{xia88,wit83}, etc), the MEM describes
the aggregation of particles with a magnetic moment and can provide useful insight into kinetic  
phenomena of great experimental and theoretical interest, such as
the growth of metallic multilayers \cite{bov98} and thin films interacting with a substrate \cite{kul06}, 
fluid adsorption on wedges \cite{rej99}, filling of templates imprinted with 
nanometer/micrometer-sized features \cite{jos01,dev03}, etc. 
Moreover, interpreting the spin of the particles in a more general way, it can represent different atomic species  
in a binary alloy, impurities or defects in a growing crystal, the states of bacteria cells like 
Salmonella \cite{sil83}, the knowledge level of students in a classroom \cite{bor01}, etc.  

The MEM growing on a small-world network could be considered, for instance, as representing the 
opinion spreading within a social group. According to the growth rules of the MEM, which are given in 
the next Section, the opinion or decision of an individual would be affected by those of their 
acquaintances, but opinion changes (analogous to spin flips in an Ising model) would not occur.  
However, as mentioned above, other interpretations of the model could also be possible in contexts as different as 
materials science, sociology, and biology.     

This paper is organized as follows: 
in Section 2, details on the model definition and the simulation method are given; 
Section 3 is devoted to the presentation and discussion of the results, while the conclusions 
are finally stated in Section 4. 

\section{The model and the simulation method}
 
In this work, we consider the one-dimensional, nearest-neighbor, adding-type 
small-world network model \cite{bol88,new99b}. Starting with a ring of $N$ 
sites and $N$ bonds, a network realization is built by adding new links connecting pairs of 
randomly chosen sites.  
For each bond in the original lattice, a shortcut is added with probability $p$. 
During this process, multiple connections between any pair of sites are avoided, as well as  
connections of a site to itself. Since the original lattice bonds are not rewired, 
the resulting network remains always connected in a single component.  
On average, $pN$ shortcuts are added and the mean coordination number is $\langle z\rangle =2(1+p)$. 
Note that $p$ can also be regarded as the mean shortcut fraction relative to the number of 
fixed lattice bonds.    

Once the network is created, a randomly chosen up or down spin is deposited on 
a random site. Then, the growth takes place by adding, one by one, further spins to the immediate neighborhood 
(the perimeter) of the growing cluster, taking into account the corresponding interaction energies. By analogy
to the Ising model, the energy $E$ of a configuration of spins is given by
\begin{equation}
E = - \frac{J}{2} \sum_
{\langle ij\rangle} S_iS_j ,
\label{energy}   
\end{equation}
where $S_i= \pm 1$ indicates the orientation of the spin for each occupied site (labeled by the 
subindex $i$), $J>0$ is the ferromagnetic 
coupling constant between nearest-neighbor (NN) spins, and 
$\langle ij\rangle$ indicates that the summation is taken over all pairs of occupied NN sites.
As with other spin systems defined on complex networks, the magnetic interaction between any pair 
of spins is only present when a network bond connects their sites. 

Setting the Boltzmann constant equal to unity ($k_{B} \equiv 1$) and measuring the absolute temperature 
$T$ in units of $J$, the probability for a new spin to be added to the (already grown) cluster is
defined as proportional to the Boltzmann factor exp$(-\Delta E /T)$, 
where $\Delta E$ is the total energy change involved. 
At each step, all perimeter sites have to be considered and the probabilities of adding a new (either up 
or down) spin to each site must be evaluated. 
Using the Monte Carlo simulation method, all growth probabilities are first computed and normalized, and then    
the growing site and the orientation of the new spin are both determined by means of a pseudo-random number.
Although the configuration energy of a MEM cluster, given by Eq.(\ref{energy}), resembles the Ising Hamiltonian, it 
should be noticed that the MEM is a nonequilibrium model in which new spins are 
continuously added, while older spins remain frozen and are not 
allowed to flip. 
The growth process naturally stops after the deposition of $N$ particles, 
when the network becomes completely filled. 

For any given set of defining parameters (i.e. the network size $N$, the shortcut-adding  
probability $p$ and the temperature $T$), ensemble averages were calculated 
over $10^4$ different (randomly generated) networks, and considering 
typically $50$ different (randomly chosen) seeds for each network configuration. 
Since all normalized growth probabilities have to be recalculated at each deposition step, 
the resulting update algorithm is rather slow. Involving a considerable computational 
effort, this work presents extensive Monte Carlo simulations that cover the whole 
shortcut-adding probability range $0\leq p\leq 1$
for different network sizes up to $N=10^4$.  
 
\section{Results and discussion}

The natural order parameter of a magnetic system is the total magnetization per site, i.e.
\begin{equation}
M={\frac{1}{N}}\sum S_i,
\end{equation}
which, in the context of this work, is to be measured on the completely filled network.
However, since MEM clusters are grown from randomly chosen seeds, the ensemble 
average of the total magnetization is $\langle M\rangle\simeq 0$. Here, we will instead 
consider the absolute value of the total magnetization, $|M|$, as the order parameter, 
which is also appropriate to avoid spurious effects arising from finite-size spontaneous 
magnetization reversals \cite{bin00,bin02}.

\begin{figure}[pt]
\begin{center}
\epsfig{figure=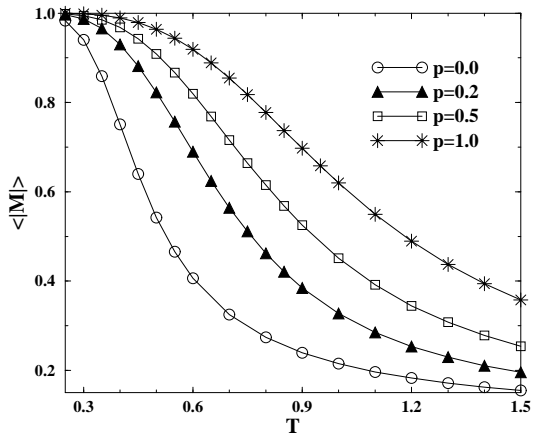,width=7.0cm}
\caption{Thermal dependence of the order parameter, for a fixed network size ($N=100$) and 
different values of the shortcut-adding probability $p$, as indicated.}
\label{fig1}
\end{center}
\end{figure}

Figure 1 shows plots of $\langle |M|\rangle$ as a function of $T$, for different values of $p$ and a fixed 
network size, $N=100$. 
The effect of increasing $p$ at a fixed temperature is that of increasing the net magnetization 
(and, hence, the order) of the system. Indeed, larger shortcut fractions favor long-range 
ordering connections between distant clusters across the network. Considering instead a 
fixed value of $p$, we see that, at low temperatures, the system grows ordered and the (absolute) magnetization
is close to unity, while at higher temperatures the disorder sets on and the magnetization  
becomes reduced significantly. 
However, fluctuations due to the finite network size prevent the magnetization from 
becoming strictly zero above the critical temperature, and the transition between the low-temperature 
ordered phase and the high-temperature disordered one becomes smoothed and rounded. 

Strictly speaking, 
Figure 1 is just showing evidence of pseudo-phase transitions, which might be precursors of true phase 
transitions taking place in the ($N\to\infty$) thermodynamic limit. 
In the following, we will proceed to characterize in more detail this pseudo-critical state, by 
measuring other observables on finite-size systems. Further on, we will use standard finite-size 
scaling procedures to establish the phase diagram $T_c$ vs $p$ corresponding to the true phase 
transition in the thermodynamic limit, as well as to calculate critical exponents that describe the 
behavior of the system at criticality. 

Let us now consider the magnetic susceptibility $\chi$, given by
\begin{equation}
\chi = \frac{N}{T}\left(\langle M^2\rangle-\langle|M|\rangle^2\right).
\label{chimi}
\end{equation}
For equilibrium systems, the susceptibility is related to order parameter fluctuations 
by the fluctuation-dissipation theorem. Although the validity of 
a fluctuation-dissipation relation in the case of a nonequilibrium system
is less evident, we will assume Eq.(\ref{chimi}) to hold also for the MEM. Indeed,  
this definition of $\chi$ proves very useful for exploring the critical behavior of this 
system, as shown in earlier studies of the MEM in regular lattices \cite{can01}, as 
well as in other nonequilibrium spin models \cite{sid98,kor01}.

\begin{figure}[pt]
\begin{center}
\epsfig{figure=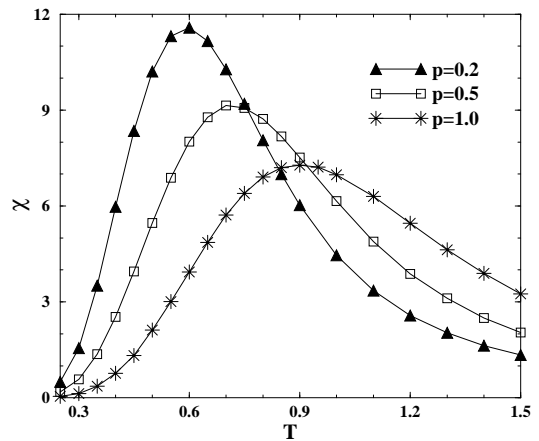,width=7.0cm}
\caption{Susceptibility as a function of the temperature, for  
different values of the shortcut-adding probability $p$ and the fixed network size $N=100$.}
\label{fig2}
\end{center}
\end{figure}

Figure 2 shows plots of $\chi$ vs $T$, for different values of $p$ and a fixed 
network size, $N=100$.
As with the thermal dependence of the order parameter shown in Figure 1, the order-disorder transitions
signaled by the peaks of the susceptibility become rounded and shifted. 
For a given probability $p$ and system size $N$, we will define the effective finite-size ``critical'' 
temperature $T_c^{eff}(N;p)$ as the temperature corresponding to the peak of the susceptibility.
Although the transition temperature of finite systems is not uniquely and precisely defined, 
the susceptibility peaks become sharper as one considers larger systems, and 
the effective finite-size ``critical'' temperatures tend to the true 
critical temperature in the (infinite-size) thermodynamic limit \cite{can01, bin00, bin02}.  

\begin{figure}[pt]
\begin{center}
\epsfig{figure=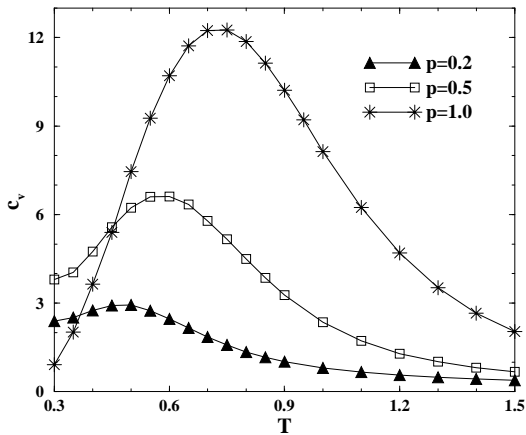,width=7.0cm}
\caption{Heat capacity per site as a function of the temperature, for a fixed network size ($N=100$) and 
different values of the shortcut-adding probability $p$, as indicated.}
\label{fig3}
\end{center}
\end{figure}

In the same vein, the heat capacity per site can be related to energy fluctuations as 
\begin{equation}
c_v =\frac{1}{NT^2}\left(\langle E^2\rangle-\langle E\rangle^2\right).
\label{heat}
\end{equation}
Figure 3 shows plots of $c_v$ vs $T$ corresponding to the same parameter values used before. 
Compared to the susceptibility, the heat capacity exhibits flatter shapes and broader peaks. 

\begin{figure}[pt]
\begin{center}
\epsfig{figure=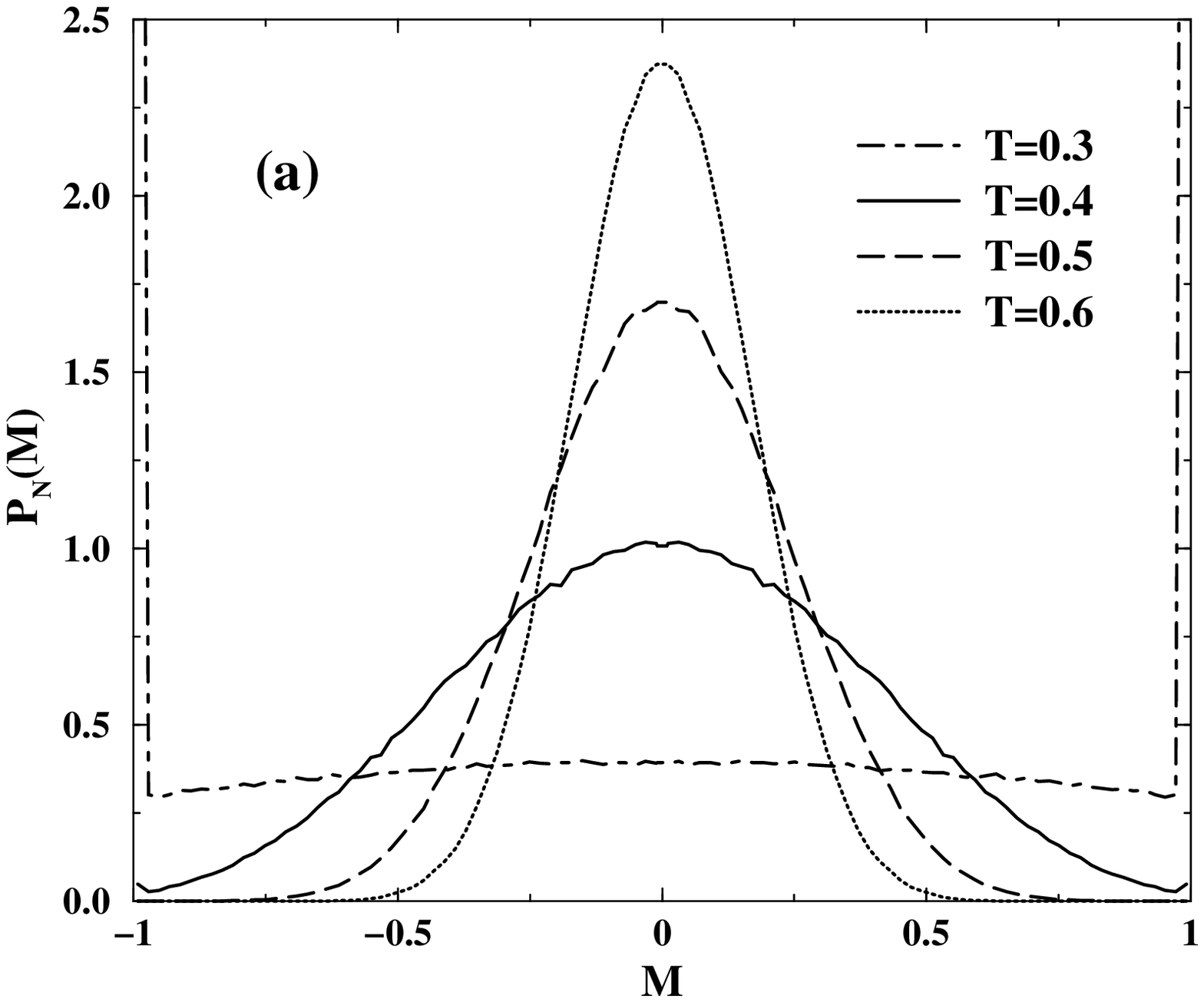,width=7.0cm}
\epsfig{figure=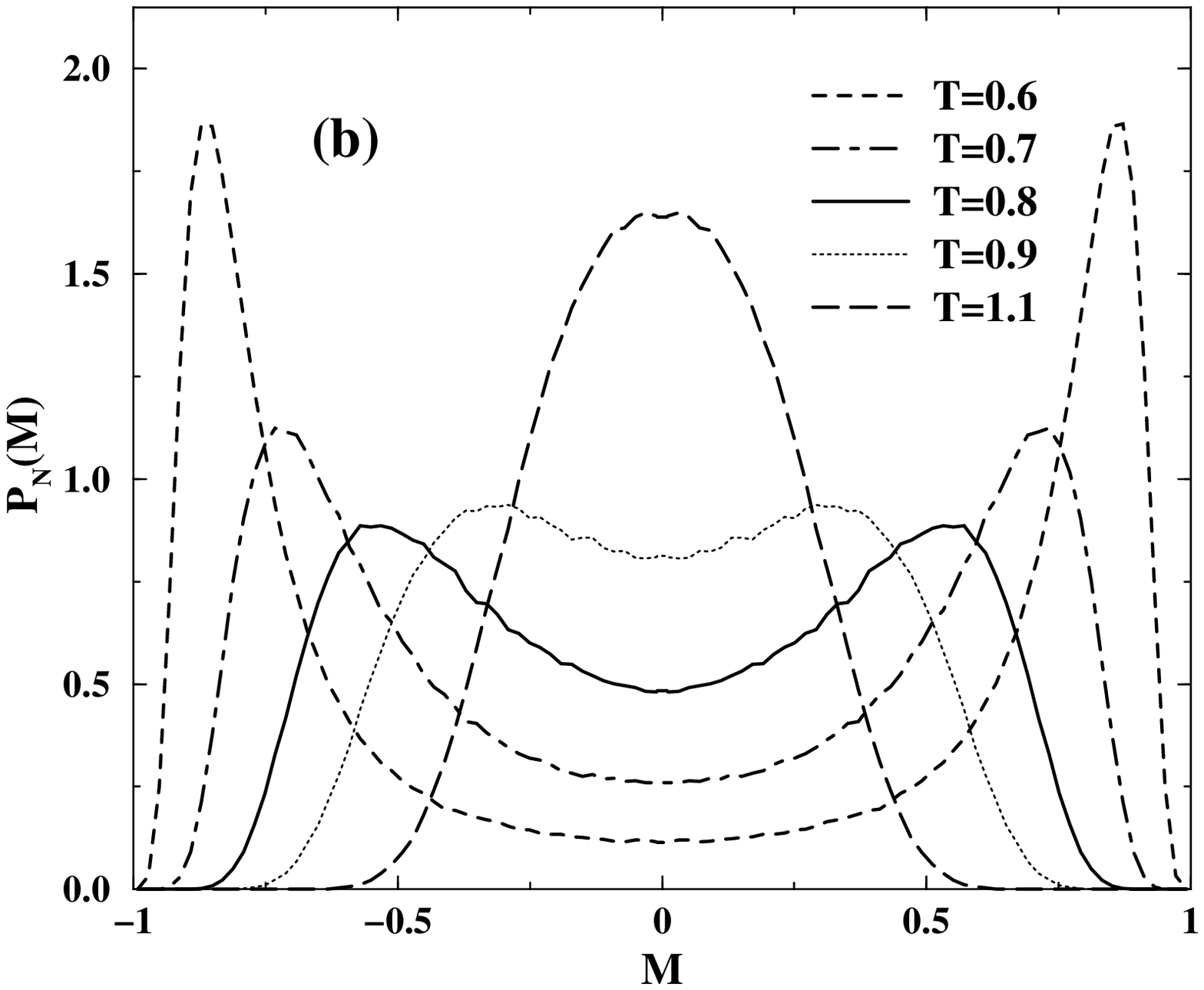,width=7.0cm}
\caption{Normalized probability distributions of the magnetization, for a fixed lattice size 
($N=1000$) and different temperatures, as indicated. (a) The ($p=0$) regular lattice, which 
exhibits local or absolute maxima at $M=0$ (the fully disordered state) and $M=\pm 1$ (the completely 
ordered states). The sharp peaks at $M=\pm 1$ for $T=0.3$ have been truncated. 
(b) The small-world network with shortcut-adding probability $p=0.5$. The gradual onset of maxima 
at $M = \pm M_{sp}$ $(0 < M_{sp} < 1)$ across the transition, 
which become sharper and approach $M= \pm 1$ as $T$ is decreased, is the hallmark of 
a true thermally-driven continuous phase transition. See more details in the text.}
\label{fig4}
\end{center}
\end{figure}

Further insight can be gained by examining the normalized probability distribution of the magnetization, $P_N(M)$. 
Fixing the network size ($N=1000$),
the behavior of $P_N(M)$ for different temperatures is shown in Figure 4 for (a) the regular lattice and
(b) the small-world network with shortcut-adding probability $p=0.5$.      
For high temperatures, in both cases the probability distributions are Gaussian-shaped and centered at $M=0$, 
as expected for thermally disordered systems. However, their behavior is quite different at 
intermediate and low temperatures.

For $p=0$ (see Figure 4(a)), the curvature of the distribution is 
convex and exhibits a local maximum always situated at $M=0$, while other maxima develop at $M=\pm 1$. 
At very low temperatures, the curved shape flattens and the maxima at $M=\pm 1$ dominate the distribution. 
Irrespective of the temperature, neither absolute nor local maxima arise at any intermediate values of 
the magnetization, i.e. different from $M=0$ (the fully disordered state) and $M=\pm 1$ (the completely 
ordered states). In fact, this noncritical behavior is in agreement with previous results for the MEM in the lattice: 
a cluster growing in a linear regular lattice from a single seed shows only a pseudo-phase 
transition with an effective ``critical'' temperature $T_c^{eff}(N)$ 
that vanishes in the ($N\to\infty$) thermodynamic limit \cite{van94}.  
An analogous behavior was observed in the MEM grown in a stripped 
(1+1)-dimensional rectangular geometry using linear seeds \cite{can01}. 

In contrast, in the small-world network (see Figure 4(b)) one observes the onset of two
maxima located at $M = \pm M_{sp}$ $(0 < M_{sp} < 1)$,
which become sharper and approach $M= \pm 1$ as $T$ is gradually decreased.
The smooth shift of the distribution maxima across $T\simeq T_c$, from $M=0$ to the low-temperature nonzero spontaneous 
magnetization $M=\pm M_{sp}$, is the signature of true thermally-driven continuous phase transitions \cite{can01,bin02}. 
Hence, critical behavior in the irreversible growth of MEM clusters arises from the presence of shortcuts in the small-world network. 
This result is a nonequilibrium realization of analogous phenomena observed in related equilibrium systems, 
as e.g. the Ising model in small-world networks generated from rewiring 1D lattices \cite{bar00,git00}. 
Notice, however, that this work focuses on nonequilibrium processes that cannot be derived from the study of equilibrium systems. 
Hence, the phenomena observed here provide further and independent evidence on the effects of long-range interactions on 
dynamical systems which are defined on small-world networks. 

\begin{figure}[pt]
\begin{center}
\epsfig{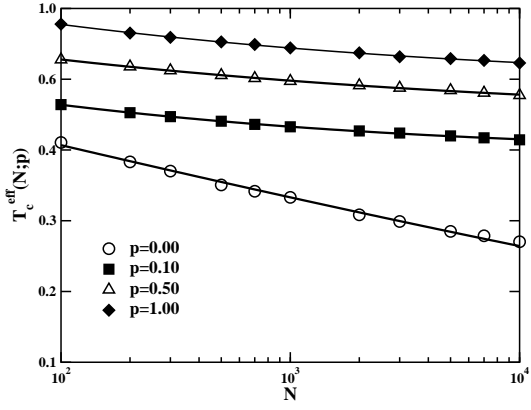}
\caption{Effective transition temperatures $T_c^{eff}(N;p)$ for $10^2\leq N\leq 10^4$ and different 
values of $p$, as indicated (symbols). Fits to the data using the finite-size scaling relation, 
Eq.(\ref{scaling1}), are also shown (solid lines). See more details in the text.} 
\label{fig5}
\end{center}
\end{figure}

In order to explore further this phenomenon, we will extrapolate the finite-size ``critical'' temperatures 
to the thermodynamic limit and build the corresponding phase diagram, in which the critical 
temperature is given as a function of the shortcut-adding probability of the network.
Moreover, this procedure will also allow to determine the critical exponents of the system. 

According to the finite-size scaling theory, developed for the treatment of finite-size effects 
at criticality and under equilibrium conditions \cite{bar83, pri90},
the difference between the true $p-$dependent 
critical temperature, $T_c(p)$, and the effective pseudo-critical one, $T_c^{eff}(N;p)$, is given by 
\begin{equation}
|T_c(p)-T_c^{eff}(N;p)|\propto N^{-1/\nu}, 
\label{scaling1}
\end{equation}
where $\nu$ is the exponent that characterizes the divergence of the correlation length 
at criticality. 

The symbols in Figure 5 show the effective transition temperatures calculated for different network sizes in the 
range $10^2\leq N\leq 10^4$ and different shortcut fractions, as indicated. Recall that, as commented above, 
the size-dependent, pseudo-critical temperatures 
$T_c^{eff}(N;p)$ were determined for different system sizes and 
shortcut fractions from the maxima of the susceptibility (see Figure 2).   
By means of least-squares fits of the finite-size scaling relation, Eq.(\ref{scaling1}), to these data one can determine 
both the true critical temperature of the system, $T_c(p)$, as well as the critical exponent $\nu$.
The non-linear least-squares fitting procedure was implemented using the Levenberg-Marquardt minimization method \cite{pre92}.  
 
The solid lines in Figure 5 show least-squares fits to the data. Indeed, 
the finite-size scaling theory proves useful in describing the behavior of this nonequilibrium   
system near the critical region, since the scaling relation given by 
Eq.(\ref{scaling1}) provides an excellent fit to all the data. 

\begin{figure}[pt]
\begin{center}
\epsfig{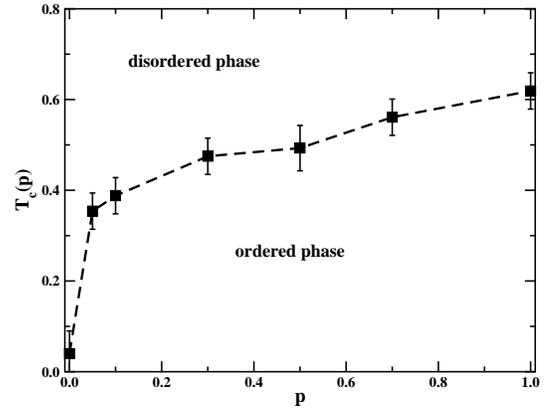}
\caption{Phase diagram $T_c(p)$ vs $p$ corresponding to the thermodynamic limit, obtained using a   
finite-size scaling relation, Eq.(\ref{scaling1}). The MEM growing on small-world 
networks with any value of $p>0$ undergoes thermally-driven continuous phase transitions. In the regular lattice 
limit, $p=0$, the system is noncritical. The dashed line is a guide to the eye.}
\label{fig6}
\end{center}
\end{figure}

Figure 6 shows the phase diagram $T_c(p)$ vs $p$, corresponding to the critical behavior of the system in 
the thermodynamic limit. The error bars reflect the statistical errors, 
which were determined from the fitting procedure. As anticipated, 
for $p>0$ the system undergoes critical order-disorder phase transitions at finite critical temperatures: 
the small-world network geometry triggers criticality. Naturally, 
the global ordering imposed by long-range shortcuts is weaker the 
lower the shortcut fraction, and hence $T_c(p)$ decreases monotonically with $p$. The critical temperature vanishes 
for $p=0$, which is the expected regular lattice limit behavior. 

As commented above, the same fits of Eq.(\ref{scaling1}) to the numerical data determine also the critical exponent $\nu$. 
The obtained ($p-$independent) value is $\nu=3.6\pm 0.4$, where the error bar reflects the statistical error resulting from the 
fitting procedure. As in the case of equilibrium spin systems defined on small-world networks (see e.g. \cite{med03,her02}), 
the universality class 
of this nonequilibrium system is not observed to depend on the shortcut density, provided that $p>0$.  

\begin{figure}[pt]
\begin{center}
\epsfig{figure=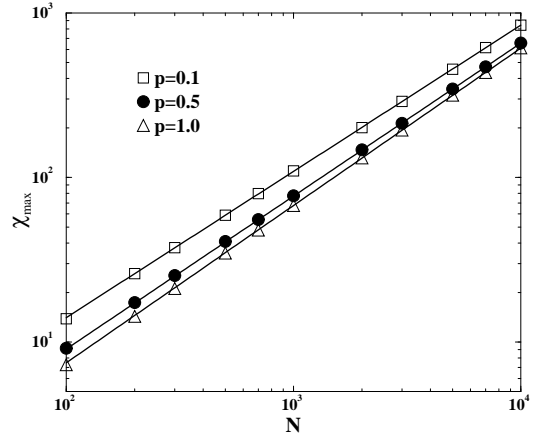,width=7.0cm}
\caption{Plots showing the maxima of $\chi$ for different network sizes in the 
range $10^2\leq N\leq 10^4$ and different shortcut fractions (symbols). Also the corresponding finite-size scaling fits, 
given by Eq.(\ref{scaling2}), are shown for comparison (solid lines).} 
\label{fig7}
\end{center}
\end{figure}

An additional characterization of the critical behavior of this system can be obtained 
by calculating the critical exponent $\gamma$, which describes the divergence of the susceptibility 
at the critical point. Using again the finite-size scaling theory \cite{bar83, pri90}, 
the exponent ratio $\gamma/\nu$ can 
be related to the peak of the susceptibility measured in finite samples of size $N$ by
\begin{equation}
\chi_{max}\propto N^{\gamma/\nu}. 
\label{scaling2}
\end{equation}
The symbols in Figure 7 correspond to the maxima of $\chi$ plotted against the network size for different values of $p$, 
as indicated, while the solid lines are fits to the data using this scaling relation, Eq.(\ref{scaling2}). 
It turns out that $\gamma/\nu=0.92\pm 0.04$, where the error bar reflects the statistical error from the fit. 
Using this ratio and the value already obtained for $\nu$, we determine $\gamma=3.3\pm 0.4$.  

\section{Conclusions}

Binary mixtures growing irreversibly on small-world networks are studied numerically by means
of extensive Monte Carlo simulations performed on the magnetic Eden model. 
Firstly, evidence for the occurrence of order-disorder pseudo-phase transitions 
is provided by the order parameter and the response functions of finite samples. 
Then, studying the order parameter distribution functions, a clearly different behavior between 
the noncritical regular lattice ($p=0$) and the small-world network ($p>0$) is observed. Indeed, the 
latter shows the behavior expected for systems undergoing thermally-driven continuous phase transitions.    
Hence, it is concluded that a small fraction of shortcuts is sufficient to trigger criticality. 

In order to obtain additional evidence of this phenomenon, standard finite-size scaling 
relations are used to determine the phase diagram $T_c$ vs $p$, in which the critical 
temperature is shown as a function of the shortcut-adding probability of the network. 
As expected, for $p>0$ the system undergoes order-disorder continuous phase transitions at 
finite critical temperatures. Since the long-range ordering is weaker the 
lower the shortcut fraction, $T_c$ decreases monotonically with $p$ and vanishes 
for $p=0$. 
Moreover, the behavior of the system at criticality is further characterized by the calculation of 
the critical exponents $\nu=3.6\pm 0.4$ and $\gamma=3.3\pm 0.4$. 

These results, obtained in the framework of nonequilibrium growth systems, are 
a novel realization of analogous phenomena, which have recently been reported in 
the investigation of related equilibrium systems. Indeed, they provide further and 
complementary evidence on the ordering 
and criticality-inducing effects of long-range interactions on 
dynamical systems defined on small-world networks.
The present work will thus hopefully stimulate and contribute to further developments 
in the fields of complex networks and nonequilibrium statistical physics. 

\acknowledgments
The author thanks Ezequiel V. Albano, Ginestra Bianconi, and Simona Rolli 
for useful discussions.


\begin{thebibliography}{100}

\bibitem{dor03} S.N. Dorogovtsev and J.F.F. Mendes, {\it Evolution of Networks}
(Oxford University Press, New York, 2003).
\bibitem{wat98} D.J. Watts and S.H. Strogatz, Nature {\bf 393}, 440 (1998). 
\bibitem{wat99} D.J. Watts, {\it Small Worlds} (Princeton University Press, Princeton, 1999).
\bibitem{new99a} M.E.J. Newman and D.J. Watts, Phys. Lett. A {\bf 263}, 341 (1999).
\bibitem{bar00} A. Barrat and M. Weigt, Eur. Phys. J. B {\bf 13}, 547 (2000).
\bibitem{git00} M. Gitterman, J. Phys. A: Math. Gen. {\bf 33}, 8373 (2000).
\bibitem{kim01} B.J. Kim, H. Hong, P. Holme, G.S. Jeon, P. Minnhagen, and M.Y. Choi, Phys. Rev. E {\bf 64}, 056135 (2001). 
\bibitem{hon02} H. Hong, B.J. Kim, and M.Y. Choi, Phys. Rev. E {\bf 66}, 018101 (2002).
\bibitem{med03} K. Medvedyeva, P. Holme, P. Minnhagen, and B.J. Kim, Phys. Rev. E {\bf 67}, 036118 (2003).
\bibitem{hin05} M. Hinczewski and A. Nihat Berker, Phys. Rev. E {\bf 73}, 066126 (2006). 
\bibitem{mon00} R. Monasson, Eur. Phys. J. B {\bf 12}, 555 (2000). 
\bibitem{jes00} S. Jespersen, I. M. Sokolov, and A. Blumen, J. Chem. Phys. {\bf 113}, 7652 (2000). 
\bibitem{sca01} A. Scala, L.A.N. Amaral, and M. Barth\'el\'emy, Europhys. Lett. {\bf 55}, 594 (2001).
\bibitem{ale02} A. Aleksiejuk, J.A. Holyst, and D. Stauffer, Physica A {\bf 310}, 260 (2002). 
\bibitem{sve02} P. Svenson and D.A. Johnston, Phys. Rev. E {\bf 65}, 036105 (2002). 
\bibitem{san02} A.D. S\'anchez, J.M. L\'opez, and M.A. Rodr\'{\i}guez, Phys. Rev. Lett. {\bf 88}, 048701 (2002).
\bibitem{her02} C.P. Herrero, Phys. Rev. E {\bf 65}, 066110 (2002).
\bibitem{kup01} M. Kuperman and G. Abramson, Phys. Rev. Lett. {\bf 86}, 2909 (2001).
\bibitem{pas01} R. Pastor-Satorras and A. Vespignani, Phys. Rev. Lett. {\bf 86}, 3200 (2001).
\bibitem{egu02} V.M. Egu\'{\i}luz and K. Klemm, Phys. Rev. Lett. {\bf 89}, 108701 (2002).
\bibitem{kle03} K. Klemm, V.M. Egu\'{\i}luz, R. Toral, and M. San Miguel, Phys. Rev. E {\bf 67}, 026120 (2003).
\bibitem{van94} N. Vandewalle and M. Ausloos, Phys. Rev. E {\bf 50}, R635 (1994).
\bibitem{can01} J. Candia and E.V. Albano, Phys. Rev. E {\bf 63}, 066127 (2001).
\bibitem{ede58} M. Eden, {\it Symposium on Information Theory in Biology}, H. P. Yockey (Ed.) (Pergamon Press, New York, 1958); 
{\it Proc. 4th Berkeley Symposium on Mathematics, Statistics and Probability}, F. Neyman (Ed.)
(University of California Press, Berkeley, 1961), Vol.IV, 223.
\bibitem{her86} H.J. Herrmann, Phys. Rep. {\bf 136}, 153 (1986). 
\bibitem{bun85} A. Bunde, H.J. Herrmann, A. Margolina, and H.E. Stanley, Phys. Rev. Lett. {\bf 55}, 653 (1985). 
\bibitem{bar94} G.C. Barker and M.J. Grimson, J. Phys. A {\bf 27}, 653 (1994). 
\bibitem{xia88} R.F. Xiao, J.I. Alexander, and F. Rosenberger, Phys. Rev. A {\bf 38}, 2447 (1988). 
\bibitem{wit83} T.A. Witten and L.M. Sander, Phys. Rev. B {\bf 27}, 5686 (1983). 
\bibitem{bov98} U. Bovensiepen, F. Wilhelm, P. Srivastava, 
P. Poulopoulos, M. Farle, A. Ney, and K. Baberschke, Phys. Rev. Lett. {\bf 81}, 2368 (1998).
\bibitem{kul06} M. Kulawik, N. Nilius, and H.-J. Freund, Phys. Rev. Lett. {\bf 96}, 036103 (2006). 
\bibitem{rej99} K. Rejmer, S. Dietrich, and M. Napi\'orkowski, Phys. Rev. E {\bf 60}, 4027 (1999).
\bibitem{jos01} D. Josell, D. Wheeler, W.H. Huber, and T.P. Moffat, Phys. Rev. Lett. {\bf 87}, 16102 (2001). 
\bibitem{dev03} A. De Virgiliis, O. Azzaroni, R.C. Salvarezza, and E.V. Albano, Appl. Phys. Lett. {\bf 82}, 1953 (2003).
\bibitem{sil83} M. Silverman and M. Simon, 
in {\it Mobile Genetic Elements}, edited by J.A. Shapiro (Academic, Orlando, 1983), p. 537.  
\bibitem{bor01} C.M. Bordogna and E.V. Albano, Phys. Rev. Lett. {\bf 87}, 118701 (2001). 
\bibitem{bol88} B. Bollob\'as and F.R.K. Chung, SIAM J. Discrete Math. {\bf 1}, 328 (1988).
\bibitem{new99b} M.E.J. Newman and D.J. Watts, Phys. Rev. E {\bf 60}, 7332 (1999).
\bibitem{bin00} D.P. Landau and K. Binder, {\it A guide to Monte Carlo simulations in 
Statistical Physics} (Cambridge University Press, Cambridge, 2000).
\bibitem{bin02} K. Binder and D.W. Heermann, {\it Monte Carlo simulation in statistical physics: 
an introduction, 4th ed.} (Springer-Verlag, Berlin, 2002).
\bibitem{sid98} S.W. Sides, P.A. Rikvold, and M.A. Novotny, Phys. Rev. Lett. {\bf 81}, 834 (1998); 
Phys. Rev. E {\bf 59}, 2710 (1999). 
\bibitem{kor01} G. Korniss, C.J. White, P.A. Rikvold, and M.A. Novotny, Phys. Rev. E {\bf 63}, 016120 (2001). 
\bibitem{bar83} M.N. Barber, {\it Phase transitions and critical phenomena}, C. Domb and J.L. Lebowitz (Eds.)  
(Academic, New York, 1983), Vol. 8.
\bibitem{pri90} V. Privman (Ed.), {\it Finite size scaling and 
numerical simulations of statistical systems} (World Scientific, Singapore, 1990).
\bibitem{pre92} W.H. Press, S.A. Teukolsky, W.T. Vetterling, and B.P. Flannery, {\it 
Numerical Recipes, 2nd Ed.} (Cambridge University Press, New York, 1992). 
\end{thebibliography}
\end{document}